\documentclass[10pt,twocolumn]{article}

\usepackage[utf8]{inputenc}
\usepackage{amsmath}
\usepackage{amsthm}
\usepackage{amssymb}
\usepackage{titlesec}
\usepackage{cite}
\usepackage{booktabs}
\usepackage{graphicx}
\usepackage[colorlinks=false]{hyperref}
\usepackage[format=plain,labelfont=it]{caption}
\usepackage[left=1.5cm,right=1.5cm,top=2cm,bottom=2cm]{geometry}
\usepackage{color}

\titleformat{\section}{\centering\normalfont\scshape}{\Roman{section}.}{5pt}{}
\titleformat{\subsection}{\normalfont\it}{\Alph{subsection}.}{5pt}{}
\titleformat{\subsubsection}{\normalfont\it}{\hspace{4mm}\arabic{subsubsection})}{5pt}{}

\newcommand\infoFootnote[1]{%
  \begingroup
  \renewcommand\thefootnote{}\footnote{#1}%
  \addtocounter{footnote}{-1}%
  \endgroup}

\newtheorem{thm}{Theorem}
\newtheorem{cor}[thm]{Corollary}
\newtheorem{lem}[thm]{Lemma}

\newtheorem{rem}{Remark}

\newcommand{\R}{\mathbb{R}} 
\newcommand{\N}{\mathbb{N}} 
 
\newcommand{\ab}{\boldsymbol{a}}

\newcommand{\cb}{\boldsymbol{c}}
\newcommand{\db}{\boldsymbol{d}}

\newcommand{\ub}{\boldsymbol{u}}
\newcommand{\vb}{\boldsymbol{v}}

\newcommand{\xb}{\boldsymbol{x}}
\newcommand{\yb}{\boldsymbol{y}}

\newcommand{\xib}{\boldsymbol{\xi}}
\newcommand{\alphab}{\boldsymbol{\alpha}}
\newcommand{\betab}{\boldsymbol{\beta}}

\newcommand{\Phib}{\boldsymbol{\Phi}}

\newcommand{\Gammab}{\boldsymbol{\Gamma}}

\newcommand{\zerob}{\boldsymbol{0}}

\newcommand{\ybs}{\mathbf{y}}
\newcommand{\ubs}{\mathbf{u}}
\newcommand{\Ab}{\boldsymbol{A}}
\newcommand{\Bb}{\boldsymbol{B}}
\newcommand{\Cb}{\boldsymbol{C}}
\newcommand{\Db}{\boldsymbol{D}}
\newcommand{\Hb}{\boldsymbol{H}}
\newcommand{\Ib}{\boldsymbol{I}}
\newcommand{\Tb}{\boldsymbol{T}}
\newcommand{\Ub}{\boldsymbol{U}}
\newcommand{\Wb}{\boldsymbol{W}}
\newcommand{\Yb}{\boldsymbol{Y}}
\newcommand{\Kb}{\boldsymbol{K}}
\newcommand{\Lb}{\boldsymbol{L}}
\newcommand{\Mb}{\boldsymbol{M}}
\newcommand{\Qb}{\boldsymbol{Q}}
\newcommand{\Rb}{\boldsymbol{R}}
\newcommand{\Fb}{\boldsymbol{F}}
\newcommand{\Eb}{\boldsymbol{E}}
\newcommand{\Gb}{\boldsymbol{G}}
\newcommand{\Vb}{\boldsymbol{V}}

\newcommand{\Mbc}{\boldsymbol{\mathcal{M}}}
\newcommand{\Obc}{\boldsymbol{\mathcal{O}}}
\newcommand{\Qbc}{\boldsymbol{\mathcal{Q}}}
\newcommand{\Rbc}{\boldsymbol{\mathcal{R}}}
\newcommand{\Hbc}{\boldsymbol{\mathcal{H}}}

\newcommand{\Tbc}{\boldsymbol{\mathcal{T}}}

\newcommand{\Vbc}{\boldsymbol{\mathcal{V}}}
\newcommand{\Uc}{\mathcal{U}}
\newcommand{\Yc}{\mathcal{Y}}

\newcommand{\Xc}{\mathcal{X}}

\newcommand{\rank}{\mathrm{rank}}

\newcommand{\blind}[1]{\textcolor{white}{#1}}
\renewcommand{\boldsymbol}[1]{#1}
\renewcommand{\mathbf}[1]{\mathrm{#1}}

\def\tvdots{\vbox{\baselineskip=2pt \lineskiplimit=0pt \kern6pt \hbox{.}\hbox{.}\hbox{.}}} 

\title{\vspace{-2mm}\bf A deterministic view on explicit data-driven (M)PC}
\author{Manuel Kl\"adtke, Dieter Teichrib, Nils Schl\"uter, and Moritz Schulze Darup\vspace{2mm}}
\date{}

\begin{document}

\maketitle

\textbf{\textit{Abstract}.} {\bf We show that the explicit realization of data-driven predictive control (DPC) for linear deterministic systems is more tractable than previously thought. To this end, we compare the optimal control problems (OCP) corresponding to deterministic DPC and classical model predictive control (MPC), specify its close relation, and systematically eliminate ambiguity inherent in DPC. As a central result, we find that the explicit solutions to these types of DPC and MPC are of exactly the same complexity. We illustrate our results with two numerical examples highlighting features of our approach.}
\infoFootnote{M. Kl\"adtke, D. Teichrib, N. Schl\"uter, and M. Schulze Darup are with the \href{https://rcs.mb.tu-dortmund.de/}{Control and~Cyberphysical Systems Group}, Faculty of Mechanical Engineering, TU Dortmund University, Germany. E-mails:  \href{mailto:manuel.klaedtke@tu-dortmund.de}{\{manuel.klaedtke, dieter.teichrib, nils.schlueter, moritz.schulzedarup\}@tu-dortmund.de}. \vspace{0.5mm}}
\infoFootnote{\hspace{-1.5mm}$^\ast$This paper is a \textbf{preprint} of a contribution to the 2022 IEEE 61st Conference on Decision and Control (CDC). The DOI of the original paper is \href{https://doi.org/10.1109/CDC51059.2022.9993384}{10.1109/CDC51059.2022.9993384}.}

\vspace{-2mm}
\section{Introduction}

\vspace{-2mm}
Data-driven predictive control (DPC), where the prediction of the systems' behavior is carried out based on collected input-output data instead of a model, is becoming more and more popular (see, e.g., \cite{Yang2013, Coulson2019DeePC,Berberich2020,Dorfler2021}). Remarkably, assuming perfect data and linear dynamics, Willems' fundamental lemma~\cite{WILLEMS2005} and variants of it (as, e.g., \cite{Waarde2020} and \cite{Markovsky2020}) allow establishing the equivalence of the data-driven and model-based approach with respect to the resulting control actions.

However, while strongly related, the two approaches lead to different optimal control problems (OCP). In fact, DPC usually results in an OCP with significantly more decision variables than model-based predictive control (MPC). As a consequence, explicit solutions of the data-driven OCP seem ``unattractive'' at first sight (especially for noisy setups \cite[Sect.~IV.B]{Alpago2020}), even for applications where explicit MPC \cite{Bemporad2002} is tractable. In fact, more decision variables typically result in significantly  more complex explicit solutions (in terms of the number of regions etc.). Yet, we show in this paper that the perceived imbalance between MPC and DPC can be completely resolved for the special case of linear deterministic systems. More precisely, we reveal that the larger number of decision variables only results in ambiguous but not more complex solutions in this case. Further, we present a simple method to systematically eliminate this ambiguity. As a central result, we obtain an explicit DPC solution of exactly the same complexity as explicit MPC.

Before detailing our approach, we briefly discuss related works from the literature. First of all, it is already well-known that the optimal input sequences resulting from deterministic DPC and MPC are identical given equivalent initial conditions \cite[Cor. 5.1]{Coulson2019DeePC}. Yet, it is also known that the original OCPs related to MPC are strictly convex while those for DPC are only convex. Thus, optimizers in DPC are typically non-unique, which significantly complicates an explicit solution. Clearly, strict convexity can be enforced through additional regularization \cite{Alpago2020,Breschi2021} (which is also helpful for noisy setups). However, this either destroys the structure we are about to identify or it renders its derivation more difficult. Alternatively, one can consider explicit DPC for fully measurable states. For this simpler case, a result similar to ours has recently been obtained in \cite{Sassella2021}. Finally, especially since we are dealing with the deterministic case and linear systems, removing ambiguity from the OCP shows many similarities to subspace identification (SID, \cite{Overschee1996}) and subspace predictive control \cite{Fiedler2021}. 
In fact, using the data matrices inherent in DPC, one could also identify a state space model and the corresponding MPC formulation would yield another equivalence. However, we provide a simple and direct approach, which can be interpreted as a tailored subspace analysis for DPC.

The remaining paper is organized as follows. In Section~\ref{sec:MPCvsDPC}, we summarize classical MPC and fundamentals of DPC. The analysis of explicit solutions of the corresponding OCPs and the central identification of a closer relation between them are carried out in Section~\ref{sec:ExplicitPC}. Finally, we illustrate our findings with two numerical examples in Section~\ref{sec:examples} and we discuss promising directions for future research in Section~\ref{sec:Conclusions}.
\vspace{-3mm}
\section{Fundamentals of MPC and DPC}
\label{sec:MPCvsDPC}
\vspace{-1mm}
\subsection{Classical MPC}
\vspace{-1mm}
We briefly summarize classical MPC in a form that is compatible with the data-driven realization in Section~\ref{subsec:dataDrivenMPC}. To this end, we assume that a linear prediction model
\begin{subequations}
\label{eq:model}
\begin{align}
    \xb(k+1)&=\Ab \xb(k)+\Bb \ub(k) \\
    \yb(k)&=\Cb \xb(k)+\Db \ub(k)
\end{align}
\end{subequations}
is known. We further assume that input and output constraints are given in terms of convex polyhedral sets  
\begin{equation}
\label{eq:constraints}
   \!\,\Uc \!:= \left\{\ub \in \R^m \left| \Mb_u \ub \leq \vb_u\!\right.\right\}\!,\,\, \Yc \!:= \left\{\yb \in \R^p \left| \Mb_y \yb \leq \vb_y\!\right.\right\}\!,\!\!\!\!\!  
\end{equation}
which are specified by the matrices $\Mb_{u/y}$ and vectors $\vb_{u/y}$, respectively. Then, classical MPC (without terminal cost and constraints) can be realized by solving
\vspace{-1mm}
\begin{align}
\label{eq:MPC}
\min_{\ub(k),\xb(k),\yb(k)} 
 &\sum_{k=0}^{N_f-1} \|\yb(k)\|_{\Qb}^2 +  \|\ub(k)\|_{\Rb}^2 \span \span \\
\nonumber
\text{s.t.} \quad \quad  \xb(0)&=\xb_0, \\
\nonumber
 \xb(k+1)&=\Ab\,\xb(k) + \Bb \ub(k), &&\forall k \in \{0,...,N_f-2\}, \\
 \nonumber
  \yb(k)&=\Cb\,\xb(k) + \Db \ub(k), &&\forall k \in \{0,...,N_f-1\}, \\
 \nonumber
\left(\yb(k),\ub(k)\right) & \in \Yc \times \Uc, &&\forall k \in \{0,...,N_f-1\}\vspace{-0.4mm}
\end{align}
in every time-step for the current state $\xb_0 \in \R^n$, where $\Qb\in\R^{p\times p}$ and $\Rb\in\R^{m\times m}$ denote weighting matrices and where $N_f \in \N$ is the prediction horizon. Now, the OCP~\eqref{eq:MPC} is typically condensed into a quadratic program (QP) such that only the inputs remain as decision variables. To this end, one first introduces the sequences
\vspace{-1mm}
\begin{equation}
     \label{eq:futureSequences}
    \ubs_f := \begin{pmatrix}
    \ub(0) \\
    \vdots \\
    \ub(N_f-1)
    \end{pmatrix} \quad \text{and} \quad \ybs_f := \begin{pmatrix}
    \yb(0) \\
    \vdots \\
    \yb(N_f-1)
    \end{pmatrix},
\end{equation}
and the augmented weighting matrices 
$\Qbc:=\mathrm{diag}(\Qb,\dots,\Qb)$ and $\Rbc:=\mathrm{diag}(\Rb,\dots,\Rb)$
to rewrite the cost function as
\begin{equation}
     \label{eq:compactCost}
    \sum_{k=0}^{N_f-1} \|\yb(k)\|_{\Qb}^2 +  \|\ub(k)\|_{\Rb}^2 = \| \ybs_f \|_{\Qbc}^2 + \| \ubs_f \|_{\Rbc}^2.
\end{equation}
We further define the matrices
\vspace{-1mm}
$$
    \Obc_N\!:=\!\begin{pmatrix}
    \Cb \\
    \Cb \Ab \\
    \vdots \\
    \Cb \Ab^{N-1}
    \end{pmatrix}
    \text{and}\,\,
     \Tbc_N\!:=\!\begin{pmatrix}
     \Db  & &  & \zerob\\
    \Cb \Bb   &  \!\ddots &  & \\
    \vdots &  \!\ddots &  \!\ddots &  \\
    \Cb \Ab^{N-2} \Bb  & \!\dots & \!\!\Cb \Bb & \!\Db
    \end{pmatrix}\!,
$$
which we will consider for different $N$ during this note. For $N=N_f$, we then obtain the relation
\vspace{-1mm}
\begin{equation}
 \label{eq:y-x-u}
    \ybs_f = \Obc_{N_f} \xb_0 + \Tbc_{N_f} \ubs_f.
\end{equation}
Finally, substituting \eqref{eq:y-x-u} into \eqref{eq:compactCost} and introducing
the augmented matrices $\Mbc_{u/y}:=\mathrm{diag}(\Mb_{u/y}, \dots, \Mb_{u/y})$ leads to
 \begin{align}
    \label{eq:ufQP}
    \ubs_f^\ast(\xb_0) := \arg \min_{\ubs_f}\,\, &\frac{1}{2} \ubs_f^\top \Hb \ubs_f + \xb_0^\top \Fb^\top \ubs_f \\
    \nonumber
     \text{s.t.} \quad &\Gb \ubs_f \leq \Eb \xb_0 + \db
\end{align}
with the parameter $\xb_0$ as well as
\begin{align}
    \nonumber
     \Hb&:= 2 \Tbc_{N_f}^\top \Qbc \,\Tbc_{N_f} + 2\Rbc, 
     &\Fb&:= 2 \Tbc_{N_f}^\top \Qbc \, \Obc_{N_f},\\
    \label{eq:HandF}
     \Gb\!&:=\! 
     \begin{pmatrix}
     \Mbc_{\ub}\\
     \Mbc_{\yb} \Tbc_{N_f} 
     \end{pmatrix}
     & \Eb&:= 
     \begin{pmatrix}
     \zerob\\
     -\Mbc_{\yb} \Obc_{N_f} 
     \end{pmatrix}, \\
     \nonumber
     \db&:= 
      \begin{pmatrix}
    \vb_{u}^\top & \dots & \vb_{u}^\top & \vb_{y}^\top &  \dots & \vb_{y}^\top 
     \end{pmatrix}^\top. \span \span
\end{align}
Remarkably, $\Hb$ is positive definite, i.e., \eqref{eq:ufQP} is strictly convex, under the assumption that $\Qb$ is positive semi-definite and that $\Rb$ is positive definite.

\subsection{DPC using input-output sequences}
\label{subsec:dataDrivenMPC}

In contrast to MPC, DPC considers input-output data instead of a model as in~\eqref{eq:model}. More precisely, DPC builds (in its simplest form) on two sequences $\ubs_d$ and $\ybs_d$ as in~\eqref{eq:futureSequences} but of length $N_d\in\N$ that reflect prerecorded system inputs and outputs. We note, at this point, that with slight abuse of notation, we denote both the elements of $\ubs_f$ and $\ubs_d$ with $\ub(k)$ (and analogously elements in $\ybs_f$ and $\ybs_d$ with $\yb(k)$). However, the specific relationship will always be clear from the context. Now, in order to realize DPC by means of $\ubs_d$ and $\ybs_d$, the sequences have to carry enough information about the systems' dynamics. This holds, for instance, if $\ybs_d$ is consistent with a persistently exciting and sufficiently long input sequence $\ubs_d$. More specifically, for deterministic DPC as considered here, consistency means that there exists a model~\eqref{eq:model} with initial state $\xb_0\in \R^n$ such that 
\begin{equation}
\label{eq:consistency}
 \ybs_d = \Obc_{N_d} \xb_0 + \Tbc_{N_d} \ubs_d.
\end{equation}
Further, according to \cite{WILLEMS2005}, $\ubs_d$ is persistently exciting of order $N_e\in \N$  if the Hankel matrix
$$
    \Hbc_{N_e}(\ubs_d):=\begin{pmatrix}
    \ub(0) & \ub(1) & \dots & \ub(N_d-N_e) \\
    \ub(1) & \ub(2) & \dots & \ub(N_d-N_e+1) \\
    \vdots & & \ddots & \vdots \\
    \ub(N_e-1) &  \ub(N_e) &\dots & \ub(N_d-1) 
    \end{pmatrix}
$$
has full row rank, i.e., $\rank(\Hbc_{N_e}(\ubs_d)) = m N_e$. This requires $\Hbc_{N_e}(\ubs_d)$ to have as least as many columns as rows, i.e., 
\begin{equation}
    \label{eq:conditionNd}
      N_d-N_e+1\geq m N_e
      \quad \Longleftrightarrow \quad N_d\geq (m+1)N_e-1.
\end{equation}
Finally, Willems' fundamental lemma \cite{WILLEMS2005} allows 
associating the given sequences $\ubs_d$ and $\ybs_d$ with other input-output sequences of the same system. In fact, under the assumption that the underlying system is linear, controllable, and $\ubs_d$ is persistently exciting of order $N_e := N_c+n$, candidate sequences $(\ubs_c,\ybs_c)$ of length $N_c \in \N$ belong to the same system as $(\ubs_d,\ybs_d)$ if and only if 
$$
    \begin{pmatrix}
         \ubs_c \\
         \ybs_c
    \end{pmatrix}
    \in \mathrm{im}\begin{pmatrix}
         \Hbc_{N_c}(\ubs_d) \\
         \Hbc_{N_c}(\ybs_d)
    \end{pmatrix}.
$$
At this point, we briefly note that recent extensions of the fundamental lemma in \cite{Waarde2020} and \cite{Markovsky2020} allow alleviating some of the restrictions above. Now, in order to utilize the previous results for DPC, we proceed similarly to \cite{Coulson2019DeePC}. We choose an integer $N_p$ equal to (or larger than) the observability index, i.e., such that the corresponding matrix  $\Obc_{N_p}$ has full column rank (which obviously requires observability). Next, we assume that $u_d$ is persistently exciting of order 
\begin{equation}
    \label{eq:choiceNe}
    N_e := N_p+N_f+n.
\end{equation}
According to the fundamental lemma, we then find that the concatenated sequences
$(\ubs_p^\top \,\,\,\ubs_f^\top)^\top$ and $(\ybs_p^\top \,\,\,\ybs_f^\top)^\top$
with 
$$
    \ubs_p := \begin{pmatrix}
    \ub(-N_p) \\
    \vdots \\
    \ub(-1)
    \end{pmatrix} \quad \text{and} \quad \ybs_p := \begin{pmatrix}
    \yb(-N_p) \\
    \vdots \\
    \yb(-1)
    \end{pmatrix}
$$
and with $(\ubs_f,\ybs_f)$ as in~\eqref{eq:futureSequences}, belong to the same system as $(\ubs_d,\ybs_d)$ if and only if
\begin{equation}
    \label{eq:uypfHa}
     \begin{pmatrix}
         \ubs_p \\
         \ubs_f \\
         \ybs_p \\
         \ybs_f
    \end{pmatrix}
    =\begin{pmatrix}
         \Hbc_{N_p+N_f}(\ubs_d) \\
         \Hbc_{N_p+N_f}(\ybs_d)
    \end{pmatrix} \ab.
\end{equation}
for some $\ab \in \R^{l}$ with $l:=N_d-N_f-N_p+1$. Based on reordering and partitioning, \eqref{eq:uypfHa} can be rewritten as 
\begin{equation}
\label{eq:blocksWpUfYf}
     \xib:= \begin{pmatrix}
         \ubs_p \\
         \ybs_p 
    \end{pmatrix} = \Wb_p \ab, \quad \ubs_f = \Ub_f \ab , \quad \text{and} \quad \ybs_f = \Yb_f \ab
\end{equation}
with the matrices $W_p$, $U_f$, and $Y_f$ representing blocks of the concatenated Hankel matrices. We are now ready to formulate the OCP associated with DPC. In fact, the combination of~\eqref{eq:compactCost} and~\eqref{eq:blocksWpUfYf} allow expressing the costs
$$
    \| \ybs_f \|_{\Qbc}^2 + \| \ubs_f \|_{\Rbc}^2 = \| \ab \|_{\Yb_f^\top \Qbc \Yb_f+\Ub_f^\top \Rbc \Ub_f}^2
$$
as a function of $\ab$. Taking into account the constraints $\Mbc_u \ubs_f \leq \Vbc_u$ and  $\Mbc_y \ubs_f \leq \Vbc_y$ and the remaining condition $\xib= \Wb_p \ab$ then leads to the QP
\begin{equation}
    \label{eq:aQP}
    \ab^\ast(\xib) := \arg \min_{\ab}\,\frac{1}{2} \ab^\top \tilde{\Hb} \ab \,\,\,\,\,\,  \text{s.t.}  \, \,\,\,\, \tilde{\Gb} \ab \leq \db,  \,\,\,\, \Wb_p \ab = \xib
\end{equation}
with the parameter $\xib$, the vector $\db$ as in~\eqref{eq:ufQP}, and  
$$
 \tilde{\Hb}:= 2 \Yb_f^\top \Qbc \Yb_f + 2\Ub_f^\top  \Rbc \Ub_f, \qquad  \tilde{\Gb}:= \begin{pmatrix}
 \Mbc_u \Ub_f \\
 \Mbc_y \Yb_f 
 \end{pmatrix}.
$$
Remarkably, the role of the initial state $\xb_0$ in \eqref{eq:ufQP} is replaced by $\xib$, i.e., the $N_p$ previous inputs and outputs, in~\eqref{eq:aQP}. Furthermore, $\ab^\ast(\xib)$ only reflects an intermediate result that is used to compute optimal inputs via ${\ubs_f^\ast(\xib):=\Ub_f \ab^\ast(\xib)}$.

\section{From explicit MPC to explicit DPC}
\label{sec:ExplicitPC}

The QP \eqref{eq:ufQP} or \eqref{eq:aQP} is typically solved for the current state~$\xb_0$ or the most recent sequences $\xib$, respectively, to obtain the optimal input for the current time-step. Subsequently, the procedure is repeated at the next sampling instance. Alternatively, in order to avoid numerical optimization during runtime, \eqref{eq:ufQP} can also be solved explicitly using parametric optimization. As a result, we then find the continuous and piecewise affine (PWA) solution
\begin{equation}
    \label{eq:ufExplicit}
    \ubs_f^\ast(\xb_0) = \left\{ \begin{array}{cc}
    \Lb_1 \xb_0 + \cb_1     &   \text{if} \quad  \xb_0 \in \Xc_1, \\
    \vdots & \vdots \\
    \Lb_s \xb_0 + \cb_s     &   \text{if} \quad  \xb_0 \in \Xc_s, 
    \end{array}\right. 
\end{equation}
which is defined on a polyhedral partition $\{\Xc_i\}_{i=1}^s$ of the state space \cite{Bemporad2002}. Computing this solution offline and evaluating it online is referred to as explicit MPC. While conceptually attractive, explicit MPC can usually only be applied for moderate ``sizes'' of the underlying QP since it is well-known that the number of regions $s\in\N$ typically grows exponentially with the number of decisions variables and constraints. As a consequence, solving \eqref{eq:aQP} parametrically seems unattractive at first sight, since especially the number of decision variables is significantly larger than in~\eqref{eq:ufQP}. In fact, while $\ubs_f$ is of dimension $m N_f$, the dimension $l$ of $\ab$ is lower-bounded by
\begin{align}
\nonumber
    l &\geq (m+1) (N_p+N_f+n)  -N_f-N_p \\
    \label{eq:lLowerBound}
     &=  m N_p+ m N_f+(m+1)n 
\end{align}
according to \eqref{eq:conditionNd} and \eqref{eq:choiceNe}. Now, while the difference of at least $m N_p+(m+1)n$ decisions variables is significant especially for $m>1$, we claim that this increase does not result in a more complex explicit solution for the special case of deterministic DPC. In fact, we show that the increase in decision variables only leads to ambiguous solutions and that this ambiguity can be removed by systematically eliminating variables using tools inspired from SID. Remarkably, simultaneously to our work, \cite{Breschi2022} proposed a conceptually similar way of eliminating decision variables for non-deterministic systems. The focus in  \cite{Breschi2022} is, however, not on explicit DPC.

\subsection{Eliminating equality constraints for DPC}

Following this claim, we initially eliminate the equality constraints in~\eqref{eq:aQP}. To this end, we assume that a generalized inverse $\Wb_p^+$  of $\Wb_p$ (satisfying the Penrose conditions) and a matrix $\Vb_{p}$ characterizing the null-space of $\Wb_p$ (i.e., $\mathrm{im}(\Vb_{p})=\mathrm{ker}(\Wb_p)$) are known. 
Then, we can substitute $\ab$ in~\eqref{eq:aQP} with 
\vspace{-1mm}
\begin{equation}
    \label{eq:aParametrized}
    \ab := \Wb_p^+ \xib + \Vb_{p} \alphab,
\end{equation}
where $\alphab$ is of dimension 
\vspace{-1mm}
\begin{equation}
    \label{eq:nu}
\nu:=\mathrm{nullity}(\Wb_p)=l-\mathrm{rank}(\Wb_p). 
\end{equation}
Clearly, the equality constraints in~\eqref{eq:aQP} are satisfied for every $\alphab \in \R^\nu$. Hence, we obtain the transformed QP 
\begin{equation}
    \label{eq:alphaQP}
    \!\alphab^\ast(\xib) = \arg \min_{\alphab} \frac{1}{2} \alphab^\top \!\hat{\Hb} \alphab + \xib^\top \! \hat{\Fb}^\top \!\alphab \,\,\,\,\text{s.t.} \,\,\,\hat{\Gb} \alphab \leq \hat{\Eb} \xib + \db\!\!\!\!
\end{equation}
with $\hat{\Hb}:=\Vb_p^\top \tilde{\Hb}\Vb_p$, $\hat{\Gb}:=\tilde{\Gb} \Vb_p$, $\hat{\Fb}:=\Vb_p^\top \tilde{\Hb} \Wb_p^+$, and ${\hat{\Eb}:=-\tilde{\Gb} \Wb_p^+}$. While the elimination of equality constraints is a standard procedure often performed internally by QP solvers, it has a useful interpretation in the case of DPC. Unlike $\ab$, which parametrizes all possible system trajectories of lengths $N_p+N_f$, the new variable $\alphab$ only parametrizes those trajectories that are consistent with $\xib$, i.e., the $N_p$ most recent inputs and outputs. However, it is important to note that \eqref{eq:aQP} is only feasible for $\xib$ belonging to the system while \eqref{eq:alphaQP} may also be feasible for other $\xib$. This observation will be relevant further below for Theorem~\ref{thm:linkKKT}.

Now, according to~\eqref{eq:nu}, the reduction of decision variables when replacing  \eqref{eq:aQP} with \eqref{eq:alphaQP} is determined by $\mathrm{rank}(\Wb_p)$. Taking into account that $\Wb_p$ contains $m N_p$ rows of the full rank matrix $\Hbc_{N_p+N_f}(\ubs_d)$, we immediately find ${\mathrm{rank}(\Wb_p)\geq m N_p}$. A closer investigation reveals the following specification.

\begin{lem} 
    \label{lem:rankWp}
    Let $\ubs_d$ and $\ybs_d$ be as in Section~\ref{subsec:dataDrivenMPC} and consider the partitions~\eqref{eq:blocksWpUfYf} of the Hankel matrices in~\eqref{eq:uypfHa}. Then,  ${\mathrm{rank}(\Wb_p)=m N_p+n}$.
\end{lem}

\begin{proof}
    It is easy to see that $\Wb_p$ can be written as
    \vspace{-1mm}
    $$
    \Wb_p = \begin{pmatrix}
             \Hbc_{N_p}(\hat{\ubs}_d) \\
             \Hbc_{N_p}(\hat{\ybs}_d)
        \end{pmatrix},
    $$
    where $\hat{\ubs}_d$ and $\hat{\ybs}_d$ refer to the sequences $\ubs_d$ respectively $\ybs_d$ shortened by the last $N_f$ elements. It is further straightforward to show that $\hat{\ubs}_d$ is persistently exciting of order $N_p+n$ (i.e., the order of $\ubs_d$ likewise reduced by $N_f$). As summarized in \cite[Sect.~I.]{Markovsky2020}, Willems' fundamental lemma \cite{WILLEMS2005} then implies $\mathrm{rank}(\Wb_p) = m N_p + n$.
\end{proof}

The combination of~\eqref{eq:lLowerBound}, \eqref{eq:nu}, and Lemma~\ref{lem:rankWp} implies
\vspace{-1mm}
\begin{equation}
    \label{eq:nuLowerBound}
\nu = l - \mathrm{rank}(\Wb_p) \geq m (N_f+n).
\end{equation}
In other words, while the number of decision variables is significantly reduced from \eqref{eq:aQP} to \eqref{eq:alphaQP}, we still find at least $m n$ more decision variables in \eqref{eq:alphaQP} than in \eqref{eq:ufQP}. Fortunately, this deficit can be eliminated as follows.

\subsection{Eliminating solution candidates in irrelevant null-spaces}

As noted in Section \ref{subsec:dataDrivenMPC}, also when applying DPC, we are mainly interested in the optimal control sequence 
\begin{equation}
    \label{eq:ufByAlpha}
     \ubs_f^\ast(\xib)=\Ub_f \ab^\ast(\xib)=\Ub_f \Wb_p^+ \xib + \Ub_f\Vb_p \alphab^\ast(\xib)
\end{equation}
(or even only in its first element). As apparent from \eqref{eq:ufByAlpha}, components of $\alphab(\xib)$ in the null-space of $\Ub_f\Vb_p$ will not affect the resulting sequence $\ubs_f(\xib)$. As a consequence, it seems promising to parametrize $\alphab$ by
\begin{equation}
    \label{eq:alphaParametrized}
    \alphab := \Kb_{f} \betab + \Vb_{f} \betab_0 ,
\end{equation}
where $\Kb_f \in \R^{\nu \times \mu}$ and $\Vb_f \!\in \R^{\nu \times \nu-\mu}$ with  $\mu :=\! \mathrm{rank}(\Ub_f\Vb_p)$ are such that $\mathrm{im}(\Vb_{f})\!=\mathrm{ker}(\Ub_f \Vb_p)$ and ${\mathrm{rank}\big((\Kb_f \,\,\,\Vb_f)\big)=\nu}$. Clearly, the columns of $\Kb_f$ and $\Vb_f$ span the subspaces that are relevant and irrelevant for $\ubs_f$, respectively. By construction, we thus obtain
\begin{equation}
    \label{eq:UfBeta0}
    \Ub_f\Vb_p \Vb_{f} \betab_0 =\zerob \quad \text{for every} \quad {\betab_0 \in \R^{\nu-\mu}}.
\end{equation} 
Hence, $\betab_0$ has no effect on the resulting input sequence $\ubs_f$. Further, since $\xib$ determines $\xb_0$ in~\eqref{eq:y-x-u} and since $\ybs_f$ is then determined by~$\ubs_f$, also $\ybs_f$ should be independent of $\betab_0$. In order to verify this hypothesis, we initially note that $\xib$ and the assumed observability allow reconstructing $\xb(-N_p)$. This state in combination with $\ubs_p$ determines~$\xb_0$. The relation is formally captured by $\xb_0 = \Gammab \xib$, where
$$
    \Gammab:= \begin{pmatrix} 
    \begin{pmatrix}
    \Ab^{N_p-1} \Bb & \dots &  \Bb
    \end{pmatrix} \!-  \Ab^{N_p} \Obc_{N_p}^+ \Tbc_{N_p} & \quad
    \Ab^{N_p} \Obc_{N_p}^+
    \end{pmatrix} 
$$
with $\Obc_{N_p}^+:=(\Obc_{N_p}^\top \Obc_{N_p})^{-1} \Obc_{N_p}^\top$.
Using this relation in~\eqref{eq:y-x-u} leads to 
\vspace{-1mm}
\begin{equation}
     \label{eq:y-xi-u}
    \ybs_f = \Obc_{N_f} \Gammab \xib + \Tbc_{N_f} \ubs_f.
\end{equation}
This equation provides the basis for a useful relation between $\Wb_p$, $\Ub_f$, and $\Yb_f$. In fact, noting that the columns of these matrices can be interpreted as uniformly shifted sequences $\xib$, $\ubs_f$, and $\ybs_f$, respectively, one finds
 \vspace{-1mm}
\begin{equation}
     \label{eq:Y-X-U}
    \Yb_f =  \Obc_{N_f} \Gammab \Wb_p + \Tbc_{N_f}  \Ub_f
\end{equation}
 as also pointed out in \cite[p.~41]{Overschee1996}. Based on this relation, we can easily derive the analogue to~\eqref{eq:UfBeta0} for output sequences.

\begin{lem} 
    Let $\Yb_f$, $\Vb_p$ and $\Vb_f$ be defined as in~\eqref{eq:blocksWpUfYf}, \eqref{eq:aParametrized} and \eqref{eq:alphaParametrized}, respectively. 
    Then,
    \vspace{-1mm}
    \begin{equation}
        \label{eq:YfBeta0}
        \Yb_f\Vb_p \Vb_{f} \betab_0  =\zerob \quad \text{for every} \quad {\betab_0 \in \R^{\nu-\mu}}.
    \end{equation} 
\end{lem}

\vspace{1mm}
\begin{proof}
    To prove the claim, we multiply \eqref{eq:Y-X-U} with $\Vb_p$ as in~\eqref{eq:aParametrized} from the right and  obtain  
    \begin{equation}
         \label{eq:YfVp}
        \Yb_f \Vb_p = \Tbc_{N_f} \Ub_f \Vb_p
    \end{equation} 
    due to $\Wb_p \Vb_p = \zerob$. Substituting~\eqref{eq:YfVp} in~\eqref{eq:YfBeta0} and taking~\eqref{eq:UfBeta0} into account completes the proof. 
\end{proof}

The relations \eqref{eq:UfBeta0} and \eqref{eq:YfBeta0} formally show that $\betab_0$ neither affects input nor output sequences parametrized by $\alphab$ as in~\eqref{eq:alphaParametrized}. As a consequence, \eqref{eq:alphaQP} can be replaced by a QP, where only $\betab \in \R^\mu$ appears as a decision variable. This central observation is formalized in the following theorem.

\begin{thm}
    Let $\Ub_f$, $\Wb_p^+$, $\Vb_p$ and $\Kb_f$ be defined as in~\eqref{eq:blocksWpUfYf}, \eqref{eq:aParametrized} and \eqref{eq:alphaParametrized}, respectively. 
    Then, the relation
    \begin{equation}
         \label{eq:ufOptViaBeta}
        \ubs_f^\ast(\xib)=\Ub_f \Wb_p^+ \xib + \Ub_f\Vb_p \Kb_f \betab^\ast(\xib)
    \end{equation} 
    \vspace{-1mm}
    holds, where 
    \vspace{-1mm}
    \begin{equation}
        \label{eq:betaQP}
        \!\betab^\ast(\xib) := \arg \min_{\betab} \frac{1}{2} \betab^\top\! \check{\Hb} \betab + \xib^\top \!\check{\Fb}^\top \!\betab    \,\,\,  \text{s.t.} \,\,\,   \check{\Gb} \betab \leq \hat{\Eb} \xib + \db\!\!\!\!
        \vspace{-1mm}
    \end{equation} 
    with $\check\Hb := \Kb_{f}^\top \hat\Hb \Kb_{f}$, $\check\Fb := \Kb_{f}^\top \hat\Fb$, and $\check\Gb := \hat\Gb\Kb_{f}$.
\end{thm}

\vspace{2mm}
\begin{proof}
    We initially show that
    \vspace{-1mm}
    \begin{equation}
         \label{eq:HFGbeta0}
        \hat{\Hb} \Vb_f \betab_0 = \zerob, \qquad \hat{\Fb}^\top \Vb_f \betab_0 = \zerob,  \quad \text{and} \quad \hat{\Gb} \Vb_f \betab_0 = \zerob
    \end{equation}
    for every $\betab_0 \in \R^{\nu-\mu}$. To see this, we first substitute the expressions for $\hat{\Hb}$, $\hat{\Fb}$, as well as $\hat{\Gb}$ and then insert $\tilde{\Hb}$, $\tilde{\Fb}$, as well as $\tilde{\Gb}$, respectively, Doing so, we obtain
    $$
        \hat{\Hb} \Vb_f \betab_0 = \Vb_p^\top \! \tilde{\Hb}\Vb_p \Vb_f \betab_0 =2 \Vb_p^\top \!(\Yb_f^\top \!\Qbc \Yb_f + \Ub_f^\top  \Rbc \Ub_f) \Vb_p \Vb_f \betab_0 
    $$
    for the first expression in~\eqref{eq:HFGbeta0}. Clearly, this expression indeed evaluates to zero for every $\betab_0 \in \R^{\nu-\mu}$ due to \eqref{eq:UfBeta0} and \eqref{eq:YfBeta0}. Analogue observations result for the remaining expressions in~\eqref{eq:HFGbeta0}. Now, the relations in \eqref{eq:HFGbeta0} imply that the choice of $\betab_0$ neither affects the cost function nor the constraints in~\eqref{eq:alphaQP} when $\alphab$ is parametrized as in~\eqref{eq:alphaParametrized}. Hence, when applying this parametrization to~\eqref{eq:alphaQP}, we can  omit the variable $\betab_0$ (or set it to zero) and restrict our attention to the new decision variable $\betab$. Formally, this  results in the QP~\eqref{eq:betaQP}.
\end{proof}

Clearly, the number of decision variables in~\eqref{eq:betaQP} equals $\mu=\mathrm{rank}(\Ub_f\Vb_p)$. Since $\Ub_f\Vb_p$ is of dimension $m N_f \times \nu$ and since~\eqref{eq:nuLowerBound} applies, we immediately find
\begin{equation}
     \label{eq:muUpperBound}
    \mu=\mathrm{rank}(\Ub_f\Vb_p) \leq \min \{ m N_f, \nu\} = m N_f.
\end{equation}
In other words, while \eqref{eq:alphaQP} definitely contains more decision variables then \eqref{eq:ufQP} according to~\eqref{eq:nuLowerBound},   \eqref{eq:betaQP} contains at most as many decision variables as \eqref{eq:ufQP} according to~\eqref{eq:muUpperBound}. At this point, it is important to note that \eqref{eq:muUpperBound} simply reflects the dimensions of $\Ub_f\Vb_p$. Recalling that \eqref{eq:ufQP} is a strictly convex QP and that \eqref{eq:betaQP} provides equivalent solutions according to~\eqref{eq:ufOptViaBeta}, already excludes the case $\mu<m N_f$. In fact, we always have $\mu = m N_f$ according to the following lemma.

\begin{lem}
\label{lem:rankUfVp}
    Let $\Ub_f$ and $\Vb_p$ be defined as in~\eqref{eq:blocksWpUfYf} and \eqref{eq:aParametrized}, respectively.  Then,  $\mathrm{rank}(\Ub_f\Vb_p)=m N_f$.
\end{lem}

\begin{proof}
    To prove the claim, we consider~\eqref{eq:uypfHa} for the special case $\xib=\zerob$, i.e., $(\ubs_p,\ybs_p)=(\zerob,\zerob)$. As apparent from~\eqref{eq:y-xi-u}, any $\ubs_f \in \R^{mN_p}$ in combination with ${\ybs_f:=\Tbc_{N_f} \ubs_f}$ leads to consistent sequences for this case. As a consequence, there exists an $\ab \in \R^l$ such that 
       $$
       \begin{pmatrix}
        \xib \\
        \ubs_f \\
        \ybs_f
    \end{pmatrix}=\begin{pmatrix}
        \zerob \\
        \ubs_f \\
        \Tbc_{N_f} \ubs_f
    \end{pmatrix} = \begin{pmatrix}
        \Wb_p \\
        \Ub_f \\
        \Yb_f
    \end{pmatrix} \ab
    $$
    for every $\ubs_f \in \R^{mN_p}$. Since we have $\zerob = \Wb_p \ab$ by construction, every such $\ab$ can be parametrized as $\ab=\Vb_p \alphab$ for a suitable $\alphab \in \R^\nu$ according to~\eqref{eq:aParametrized}. Now, since the choice of $\ubs_f \in \R^{m N_f}$ is not restricted, we find $\mathrm{im}(\Ub_f \Vb_p)=\R^{mN_f}$, which immediately completes the proof.
\end{proof}

Before analyzing implications of $\mu = m N_f$, we briefly note that Lemma~\ref{lem:rankUfVp} also allows to specify the choice of $\Kb_f$.

\begin{lem}
\label{lem:KfCharacterization}
    Any $\Kb_f$ that complies with the parametrization in~\eqref{eq:alphaParametrized} can be written as 
    \begin{equation}
        \label{eq:KfCharacterization}
        \Kb_f = \Vb_p^\top \Ub_f^\top \Phib
    \end{equation}
    for some non-singular matrix $\Phib \in \R^{m N_f \times m N_f}$.
\end{lem}

\begin{proof}
    By construction of~\eqref{eq:alphaParametrized}, the column space of $\Kb_f$ has to be equal to the row space of $\Ub_f \Vb_p$. Since $\Ub_f \Vb_p$ has full row rank according to Lemma~\ref{lem:rankUfVp}, the row space is, for example, spanned by the $m N_f$ columns of $\Vb_p^\top \Ub_f^\top$. Clearly, any other basis of the row space can be obtained according to~\eqref{eq:KfCharacterization} by a suitable transition matrix $\Phib$.
\end{proof}

\subsection{Two sides of the same coin}

The parametrizations~\eqref{eq:aParametrized} and \eqref{eq:alphaParametrized} reveal a novel relation between MPC and deterministic DPC that goes beyond existing studies of the close relationship (as, e.g., in \cite[Sect.~V.D]{Coulson2019DeePC}).
In fact, the QP \eqref{eq:ufQP} associated with MPC is formally related to the DPC variant \eqref{eq:betaQP} as follows.

\begin{lem}
\label{lem:relationsHFGE}
The cost and constraint specifications of \eqref{eq:ufQP} and \eqref{eq:betaQP} satisfy the relations
 \begin{subequations}
 \label{eq:HFGEcheckRelatedToHFGE}
  \begin{align}
  \label{eq:HcheckRelatedToH}
        \check{\Hb}  &= \Kb_{f}^\top\Vb_p^\top \Ub_f^\top \Hb \Ub_f \Vb_p\Kb_{f}, \\
 \check{\Fb} &=  \Kb_{f}^\top\Vb_p^\top \Ub_f^\top \Fb \Gammab \Wb_p\Wb_p^+  + \Kb_{f}^\top\Vb_p^\top\Ub_f^\top \Hb \Ub_f\Wb_p^+, \\
\check{\Gb}  &=\Gb \Ub_f \Vb_p\Kb_{f},  \quad \text{and} \quad   \hat{\Eb}= \Eb \Gammab \Wb_p\Wb_p^+  - \Gb \Ub_f \Wb_p^+.  
    \end{align}
 \end{subequations}
\end{lem}

\vspace{1mm}
\begin{proof}
In order to prove \eqref{eq:HcheckRelatedToH}, we note that
\vspace{-1mm}
\begin{align*}
  \check{\Hb}  
   &= 2 \Kb_{f}^\top \Vb_p^\top  \left( \Yb_f^\top \Qbc \,\Yb_f + \Ub_f^\top  \Rbc  \Ub_f \right) \Vb_p  \Kb_{f}  \\
   & = 2 \Kb_{f}^\top \Vb_p^\top  \Ub_f^\top \Tbc_{N_f}^\top \Qbc\, \Tbc_{N_f} \Ub_f \Vb_p  + 2 \Vb_p^\top \Ub_f^\top  \Rbc  \Ub_f  \Vb_p \Kb_{f} \\
   & =\Kb_{f}^\top  \Vb_p^\top \Ub_f^\top \Hb \Ub_f \Vb_p \Kb_{f}
\end{align*}
by definition of $\check{\Hb}$, $\hat{\Hb}$ and $\tilde{\Hb}$, due to \eqref{eq:YfVp}, and by definition of $\Hb$ in \eqref{eq:HandF}, respectively. The remaining relations in~\eqref{eq:HFGEcheckRelatedToHFGE} can be proven analogously.
\end{proof}

In principle, we can state a similar result to Lemma~\ref{lem:relationsHFGE} for the relation between \eqref{eq:ufQP} and \eqref{eq:alphaQP}. However, only~\eqref{eq:HFGEcheckRelatedToHFGE} involves the terms $\Ub_f \Vb_p\Kb_{f}$ with the following useful feature.

\begin{lem}
\label{lem:UfVpKfNonSingular}
Let $\Ub_f$, $\Vb_p$, and $\Kb_f$ be defined as in~\eqref{eq:blocksWpUfYf}, \eqref{eq:aParametrized}, and \eqref{eq:alphaParametrized}, respectively. Then, $\Ub_f \Vb_p\Kb_{f}$ is non-singular.
\end{lem}

\begin{proof}
    We initially find $\Ub_f \Vb_p\Kb_{f}=\Ub_f \Vb_p \Vb_p^\top \Ub_f^\top \Phib$ for some non-singular $\Phib$ according to Lemma~\ref{lem:KfCharacterization}. Further, since $\Ub_f \Vb_p$ has full row rank by Lemma~\ref{lem:rankUfVp}, $\Ub_f \Vb_p \Vb_p^\top \Ub_f^\top$ is non-singular and, hence, also the product $\Ub_f \Vb_p \Vb_p^\top \Ub_f^\top \Phib$.
\end{proof}

Lemma~\ref{lem:UfVpKfNonSingular} immediately leads to the following major result.

\begin{lem}
The QP~\eqref{eq:betaQP} is strictly convex.
\end{lem}

\begin{proof}
Since $\Ub_f \Vb_p\Kb_{f}$ is non-singular by Lemma~\ref{lem:UfVpKfNonSingular},   $\check{\Hb}$ and $\Hb$ are congruent according to \eqref{eq:HcheckRelatedToH}. Hence, $\check{\Hb}$ inherits the positive definiteness of $\Hb$, which proves the claim.
\end{proof}

We are now ready to address the explicit solutions of \eqref{eq:ufQP} and \eqref{eq:betaQP}. To this end, we recall that~\eqref{eq:ufExplicit} can be derived from the parametric Karush-Kuhn-Tucker (KKT) conditions 
\vspace{-1mm}
\begin{subequations}
\label{eq:uOptKKT}
\begin{align}
\label{eq:uOptKKTa}
     \Hb \ubs_f^\ast(\xb_0) + \Fb\xb_0 +\Gb^\top\lambda^\ast(\xb_0) &= \zerob, \\
    \Gb\ubs_f^\ast(\xb_0) - \Eb\xb_0-\db &\leq \zerob, \\
    \lambda^\ast(\xb_0) &\geq \zerob, \\
    \mathrm{diag}\left(\lambda^\ast(\xb_0)\right) \left(\Gb\ubs_f^\ast(\xb_0)-\Eb\xb_0 - \db\right) &= \zerob
\end{align}
\end{subequations}
of \eqref{eq:ufQP} \cite[Sect.~4.1]{Bemporad2002}. 
Analogously, the explicit solution of \eqref{eq:betaQP} follows from the parametric KKT conditions
\vspace{-1mm}
\begin{subequations}
\label{eq:betaOptKKT}
\begin{align}
\label{eq:betaOptKKTa}
     \check{\Hb} \betab^\ast(\xib) + \check{\Fb} \xib +\check{\Gb}^\top \check{\lambda}^\ast(\xib) &= \zerob, \\
    \check{\Gb} \betab^\ast(\xib) - \hat{\Eb}\xib-\db &\leq \zerob, \\
   \check{\lambda}^\ast(\xib) &\geq \zerob, \\
    \mathrm{diag}\left(\check{\lambda}^\ast(\xib)\right) \big( \check{\Gb} \betab^\ast(\xib)-\hat{\Eb}\xib - \db\big) &= \zerob.
\end{align}
\end{subequations}
A central observation now is that~\eqref{eq:uOptKKT} and \eqref{eq:betaOptKKT}  are equivalent.\!

\begin{thm}\label{thm:linkKKT}
    The KKT conditions~\eqref{eq:uOptKKT} and \eqref{eq:betaOptKKT} are coupled by the relations~\eqref{eq:ufOptViaBeta}, $\xb_0 = \Gammab \Wb_p\Wb_p^+ \xib$, and $\lambda^\ast(\xb_0)=\check{\lambda}^\ast(\xib)$.
\end{thm}

\begin{proof}
Substituting the coupling relations in~\eqref{eq:uOptKKT} and multiplying~\eqref{eq:uOptKKTa} with the transpose of $\Tb:=\Ub_f \Vb_p \Kb_f$ from the left, and taking \eqref{eq:HFGEcheckRelatedToHFGE} into account, immediately allows us to transform \eqref{eq:uOptKKT} into \eqref{eq:betaOptKKT}. 
The inverse transformation follows analogously by noting that $\Tb$ is invertible according to Lemma~\ref{lem:UfVpKfNonSingular}, which, e.g., allows to derive $\beta^\ast(\xib) = \Tb^{-1} (\ubs_f^\ast(\xb_0) - \Ub_f \Wb_p^+ \xib)$ from \eqref{eq:ufOptViaBeta}.
\end{proof}
\begin{rem}
The relation between $\xb_0$ and $\xib$ has initially been introduced as $\xb_0=\Gammab \xib$ above~\eqref{eq:y-xi-u}. However, we require  $\xb_0 = \Gammab \Wb_p \Wb_p^+\xib$ in Theorem~\ref{thm:linkKKT} in order to account for $\xib$ not belonging to the system but feasible for~\eqref{eq:alphaQP}. In fact, $\Wb_p \Wb_p^+\xib$ maps such $\xib$ to belonging ones (and leaves already belonging $\xib$ unaltered). 
\end{rem}

Based on the equivalence of the KKT conditions, it is straightforward to see that also the explicit solutions of \eqref{eq:ufQP} and \eqref{eq:betaQP} are equivalent. Most importantly, we find the following result that we state without a formal proof.

\begin{cor}
\label{cor:sIsTheSame}
Assume that the explicit solution of \eqref{eq:ufQP} can be described based on a continuous PWA function with $s$ segments as in~\eqref{eq:ufExplicit}. Then, the same applies to the explicit solution of \eqref{eq:betaQP} and vice versa. 
\end{cor}
\begin{rem}
Note that a similar statement could, in principle, also be formulated for the solution $\ab^\ast(\xib)$ of \eqref{eq:aQP}. In fact, by exploiting  \cite[Cor. 5.1]{Coulson2019DeePC}, it immediately follows that $\ubs_f(\xi)=U_f \ab^\ast(\xib)$ can also be described with $s$ segments. However, even for fixed data matrices,  $\ab^\ast(\xib)$ is not unique, which significantly complicates the derivation of an explicit solution (without using the tools leading to \eqref{eq:betaQP}).
\end{rem}

\section{Numerical examples}
\label{sec:examples}

\subsection{Illustrating key insights with a $1$-dimensional system}
\label{subsec:firstExample}

\begin{figure}[tp]
        \centering
        \includegraphics{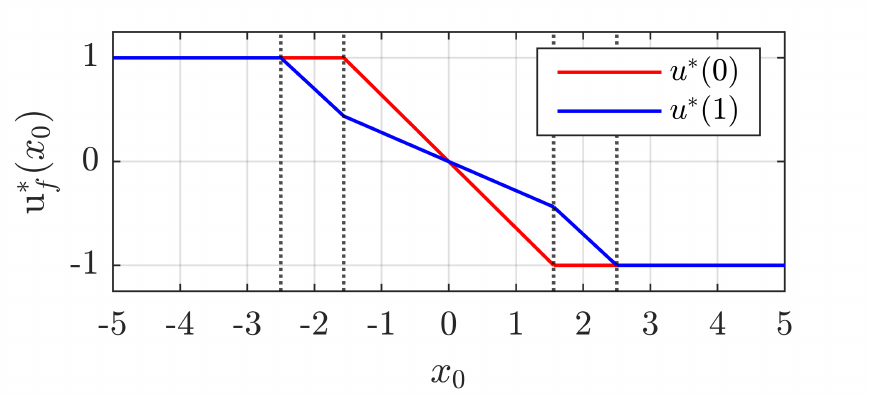}
        \vspace{-3mm}
        \caption{Explicit solution $\ubs_f^\ast(\xb_0)$ for MPC.}
        \label{fig:ufOpt}
    \end{figure}

As a first example, we consider system~\eqref{eq:model} with
\vspace{-1mm}
$$
\Ab=1.2 \quad \text{and} \quad \Bb=\Cb=\Db=1 \vspace{-1mm}
$$
subject to the constraints $\Uc=[-1,1]$ and $\Yc=[-4,4]$. Further, we choose $\Qb=\Rb=0.5$ and $N_f=2$, 
which already determines the MPC problem~\eqref{eq:MPC}. 
In order to specify~\eqref{eq:ufQP}, we note that $\Mb_u=\Mb_y=(1\,\,\,-1)^\top$, $\vb_u=(1\,\,\,1)^\top$ and $\vb_y=(4\,\,\,4)^\top$ are in line with~\eqref{eq:constraints}.
Explicitly solving~\eqref{eq:ufQP} then leads to the PWA functions in Figure~\ref{fig:ufOpt} with $s=5$ segments.

\begin{figure}[tp]
\vspace{-2mm}
    \centering
    
    \includegraphics{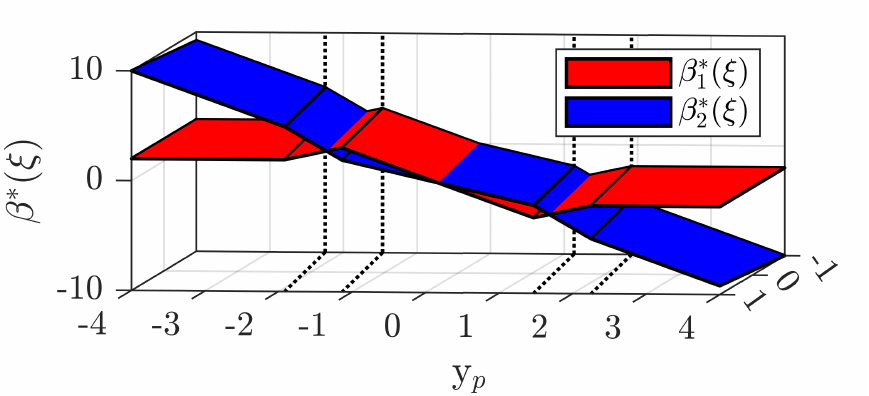}
    \vspace{-3mm}
    \caption{Explicit solution $\betab^\ast(\xib)$ for (modified) DPC. Note that $(y_p,u_p)$ is artificially restricted to $\Yc \times \Uc$ for visualization.}
    \label{fig:betaOpt}
    \vspace{-2mm}
\end{figure}

Now, to setup and investigate the DPC, we first note that $N_p=1$ guarantees full rank of $\Obc_{N_p}=\Cb=1$. Hence, we choose an input sequence $\ubs_d$, which is persistently exciting of order $N_e=4$ as in~\eqref{eq:choiceNe}. According to~\eqref{eq:conditionNd}, this requires at least $N_d=7$ elements. It can be easily verified that 
 \vspace{-1mm}
 $$
 \ubs_d:=\begin{pmatrix}
     -0.6 &  \,0 & \,0 & \,0 & \,0.5 & \,0.5 & \,1
\end{pmatrix}^\top
$$
satisfies all conditions. Furthermore, 
\vspace{-1mm}
$$
\ybs_d:=\begin{pmatrix}
     -0.1 &  0 & 0 & 0 & 0.5 & 1 & 2.1
\end{pmatrix}^\top
$$
is a consistent output sequence since~\eqref{eq:consistency} is satisfied for ${\xb_0=0.5}$. According to~\eqref{eq:blocksWpUfYf}, $\ubs_d$ and $\ybs_d$ specify
$$
    \Wb_p =\! 
    \begin{pmatrix}
        -0.6	&\!\!0	&\!\!0	&\!\!0	&\!\!0.5\\
        -0.1	&\!\!0	&\!\!0	&\!\!0	&\!\!0.5
    \end{pmatrix}\!,
\,\,\,\,
    \Ub_f = \!
    \begin{pmatrix}
        0	&\!\!0	&\!\!0	    &\!\!0.5	&\!\!0.5 \\
        0	&\!\!0	&\!\!0.5	&\!\!0.5	&\!\!1
    \end{pmatrix}\!,
$$
and $\Yb_f$ with $l=5$. In the following, we mainly focus on the transformation to~\eqref{eq:betaQP} and its explicit solution. 
To this end, we first require $\Wb_p^+$ and $\Vb_p$ as in~\eqref{eq:aParametrized}. Taking $\mathrm{rank}(\Wb_p)=2$ and, consequently, $\nu=3$ into account, suitable choices are
\vspace{-1mm}
$$
    \Wb_p^+ = 
    \begin{pmatrix}
        -2 & 2 \\
        \blind{+}0 & 0 \\
        \blind{+}0 & 0 \\
        \blind{+}0 & 0 \\
        -0.4 & 2.4
    \end{pmatrix}
    \quad \text{and} \quad
    \Vb_p = 
    \begin{pmatrix}
        0 & 0 & 0 \\
        1 & 0 & 0 \\
        0 & 1 & 0 \\
        0 & 0 & 1 \\
        0 & 0 & 0 
    \end{pmatrix}.\vspace{-1mm}
$$
We next focus on the parametrization in~\eqref{eq:alphaParametrized} and choose
\vspace{-1mm}
$$
\Kb_f^\top=2 \Ub_f \Vb_p  = \begin{pmatrix}
        0 & 0 & 1\\
        0 & 1  & 1
    \end{pmatrix}
$$
in accordance with~\eqref{eq:KfCharacterization} for $\Phib=2\Ib_2$.
This specifies~\eqref{eq:betaQP}, where we only list
$$
\check{\Hb} = 
    \begin{pmatrix}
         1.75 & \,2.5\\
       2.5 & \,3.75
    \end{pmatrix}  \quad \text{and} \quad \check{\Fb} = 
    \begin{pmatrix}
         -1.34 & \,8.04\\
       -1.96 & \,11.76
    \end{pmatrix}
$$
as a reference.  Explicitly solving~\eqref{eq:betaQP} leads to the PWA functions in Figure~\ref{fig:betaOpt}. Obviously, $\betab^\ast(\xib)$ likewise consists of $s=5$ segments as predicted by Corollary~\ref{cor:sIsTheSame}.

\subsection{Investigating practical features with the double integrator}

As a second example, we consider a standard double integrator system with 
$$
    \Ab = 
    \begin{pmatrix}
        1 & 1 \\
        0 & 1
    \end{pmatrix},\;\;
    \Bb = 
    \begin{pmatrix}
        0.5 \\
        1
    \end{pmatrix},\;\;
    \Cb = 
    \begin{pmatrix}
        1 &
        0
    \end{pmatrix}, \;\; \text{and}\;\;
    \Db = 0
$$
subject to the constraints $\Uc=[-1,1]$ and $\Yc=[-25,25]$.
Further, we choose $N_f=5$ as well as $\Qb=\Ib_2$ and $\Rb=0.01$.
We next reformulate the constraints as in the first example with $\vb_y=(25\,\,\,\,\,25)^\top$ and explicitly solve~\eqref{eq:ufQP} using the multi-parametric toolbox \cite{MPT3}. As a result, we obtain ${m N_f=5}$ PWA functions with $s=33$ segments.

The focus of the following analysis of the modified DPC is slightly different to that in Section~\ref{subsec:firstExample}.
In fact, while the first example aimed for an as simple as possible illustration of the novel approach, this second example addresses more practical implementations. More specifically, we investigate the influence of randomly chosen input sequences $\ubs_d$ with larger lengths $N_d$ than theoretically required. In this context, we initially note that full rank of $\Obc_{N_p}$ requires $N_p \geq 2$. As a consequence, we need at least $N_d \geq 17$ to achieve persistent excitation of order $N_e \geq 9$. Hence, DPC initially results in the QP \eqref{eq:aQP} with $l \geq 11$ decision variables. Next, by eliminating the equality constraints, we find \eqref{eq:alphaQP} with $\nu \geq 7$.
As indicated by \eqref{eq:muUpperBound} and Lemma~\ref{lem:rankUfVp}, the final simplification step always leads to the QP \eqref{eq:betaQP} with $\mu = m N_f = 5$ and, hence, as many decision variables as~\eqref{eq:ufQP} independent of the actual choices of $N_d$ and $N_p$.
In addition, also the number of segments of the explicit PWA solution to \eqref{eq:betaQP} is identical to that of~\eqref{eq:ufQP}. These observations can be useful in practice since lower bounds for $N_d$ and $N_p$ might not always be available.

\vspace{-0.25mm}

\section{Conclusions and Outlook}
\label{sec:Conclusions}\vspace{-0.25mm}
By establishing a stricter relation to classical MPC, we have shown that explicit DPC for deterministic linear systems is not as intractable as the ``dimensions'' of the corresponding OCP suggest. More precisely, through SID-type manipulations of the involved data matrices, we expressed DPC in terms of a strictly convex parametric QP that has exactly as many decisions variables and an exactly as complex explicit solution as its MPC counterpart.

Deterministic DPC for linear systems is of limited use for practical applications, which typically involve uncertainties and nonlinear effects. Hence, future work will address extensions to noisy and uncertain data as well as nonlinear systems. In this context, a promising direction could be the estimation of the ``deterministic part'' of the system as recently proposed in \cite{Breschi2022}.
Furthermore, we will investigate potential applications of explicit DPC such as, e.g., the extension of the encrypted DPC without constraints in \cite{Alexandru2020Towards} to a realization involving the constraints~\eqref{eq:constraints}.
\vspace{-1mm}
\newpage

\end{document}